\documentclass[prl,aps,twocolumn,epsfig,showpacs]{revtex4}
\usepackage{graphicx,epsf}
\usepackage{amsmath}
\usepackage{bm}
\usepackage{pstricks}
\usepackage{psfrag}

\newcommand{\beq}{\begin{equation}}
\newcommand{\eeq}{\end{equation}}
\newcommand{\beqn}{\begin{eqnarray}}
\newcommand{\eeqn}{\end{eqnarray}}

\newcommand{\ua}{\uparrow}
\newcommand{\da}{\downarrow}

\def\tit#1#2#3#4#5{{#1}{\bf #2}, #3 (#4)}

\def\prl{Phys.\ Rev.\ Lett.\ }

\def\prb{Phys.\ Rev.\ B\ }

\def\natu{Nature\ }

\def\bei{\begin{itemize}}
\def\eei{\end{itemize}}

\begin{document}


\title{Analysis of a SU(4) generalization of Halperin's wave functions as an
approach towards a SU(4) fractional quantum Hall effect in graphene sheets}

\author{M. O. Goerbig$^1$ and N. Regnault$^2$}

\affiliation{$^1$Laboratoire de Physique des Solides,
Univ. Paris-Sud, CNRS UMR 8502, F-91405 Orsay, France}

\affiliation{$^2$Laboratoire Pierre Aigrain, D\'epartement de Physique,
Ecole Normale Sup\'erieure,
24 rue Lhomond, F-75005 Paris, France
}

\date{\today}

\begin{abstract}

Inspired by the four-fold spin-valley symmetry of relativistic electrons in graphene,
we investigate a possible SU($4$) fractional quantum Hall effect, which may also
arise in bilayer semiconductor quantum Hall systems with small Zeeman gap. SU($4$) 
generalizations of Halperin's wave functions
[Helv. Phys. Acta {\bf 56}, 75 (1983)], which may break differently the original SU($4$) symmetry,
are studied analytically and compared, at $\nu=2/3$, to exact-diagonalization studies.

\end{abstract}


\maketitle

The fractional quantum Hall effect (FQHE) in conventional semiconductor (typically GaAs)
heterostructures may be understood to great extent with the 
help of one-component U($1$) trial wave functions, such as Laughlin's
\cite{laughlin} or composite-fermion generalizations of it \cite{jain}. The effect occurs due to the
repulsive electronic interactions when the filling factor, the ratio $\nu=n_{el}/n_B$ between the 
electronic density $n_{el}$ and the flux density 
$n_B=B/(h/e)$, belongs to the ``magic" series $\nu=p/(2sp\pm 1)$, 
with integral $s$ and $p$, or particle-hole symmetric fillings. 

Soon after Laughlin's original proposal, Halperin pointed out that the one-component picture for the
FQHE may have some shortcomings due to a relatively small Zeeman effect, $\Delta_Z\sim 0.3 B{\rm [T]}$K, 
which does not outcast in GaAs the
characteristic interaction energy $e^2/\epsilon l_B\sim 50 \sqrt{B{\rm [T]}}$K until extremely high 
magnetic fields \cite{halperin}. Instead, he proposed a two-component generalization of Laughlin's
wave function in order to account for the spin SU($2$) degree of freedom \cite{halperin}. Indeed, 
polarization measurements indicate that several states, such as e.g. $\nu=2/3$ and $2/5$, are not 
spin-polarized and may sometimes undergo transitions between competing FQHE states with
different spin polarization when the total magnetic field is varied at constant filling factor 
\cite{expPol2}.

The recent observation of a relativistic integral quantum Hall (QH) effect in graphene 
\cite{novoselov1,zhang1} has led to several theoretical investigations concerning an eventual
FQHE in this novel carbon compound \cite{castroneto,COgraphene,AC,toke,khvesh,toke2}. In contrast to
the abovementioned GaAs heterostructures, graphene has an underlying SU($4$) symmetry due to the
two-fold valley degeneracy in addition to the physical spin degree of freedom. The Coulomb interaction
respects this symmetry to lowest order in $a/l_B\sim 10^{-2}...10^{-1}$, 
where $a=0.14$nm is the distance between nearest-neighbor carbon atoms, and $l_B=\sqrt{\hbar/eB}=
25/\sqrt{B{\rm [T]}}$nm is the magnetic length
\cite{COgraphene,alicea}. In previous exact-diagonalization studies
on potential candidates of a FQHE 
in graphene \cite{AC,toke}, the SU($4$) symmetry has been omitted and a complete
spin polarization presupposed. In this case, the two-fold valley degeneracy may be mimicked by
a SU($2$) isospin, and graphene may be treated as a non-relativistic
SU($2$) QH system if one replaces the non-relativistic by the relativistic effective potentials
\cite{COgraphene,nomura}. More recently, T\"oke and Jain have proposed a composite-fermion 
construction at $\nu=p/(2sp+1)$ with an internal SU($4$) symmetry \cite{toke2}.

Here, we investigate a graphene SU($4$) FQHE, where we 
omit the assumption of complete spin-polarization
because the ratio $\Delta_Z/(e^2/\epsilon l_B)\simeq 0.003...0.014\sqrt{B{\rm [T]}}$K remains small
(the precise value depends on the
effective dielectric constant which may vary from $\epsilon \simeq 1$ for a free graphene layer to
$\epsilon \simeq 5$ on a SiO substrate \cite{alicea}). We propose a SU($4$) generalization 
of Halperin's wave function, which may lead to a plethora of new
FQHE states not captured in previous studies.
We mainly discuss the filling factor $\nu=2/3$, where our SU($4$) 
exact-diagonalization results indicate rich physics beyond 
the SU($2$) case \cite{AC,toke}. Notice that the graphene filling
factor $\nu_G$ is defined with respect to the Dirac points, and the central Landau level (LL) is therefore
half-filled at $\nu_G=0$, whereas in semiconductor QH systems $\nu$ is defined
with respect to the bottom of the lowest LL. In order to make the connection between the two of them, 
taking into account the four-fold degeneracy, one has to choose $\nu_G=-2+\nu$ (or $\nu_G=2-\nu$, due
to electron-hole symmetry). 


In the spirit of Ref. \cite{halperin}, we generalize Laughlin's
wave function \cite{laughlin} to SU($K$), with $K$ components,
\beq\label{eq01}
\psi_{m_1,...,m_K;n_{ij}}^{SU(K)}
=\phi_{m_1,...,m_K}^L \phi_{n_{ij}}^{inter} e^{-\sum_{j=1}^{K}
\sum_{k_j=1}^{N_j}|z_{k_j}^{(j)}|^2/4},
\eeq
where
$$
\phi_{m_1,...,m_K}^L=\prod_{j=1}^K\prod_{k_j<l_j}^{N_j}\left(
z_{k_j}^{(j)}-z_{l_j}^{(j)}\right)^{m_j}
$$
is a product of $K$ Laughlin wave functions, and
$$
\phi_{n_{ij}}^{inter}=\prod_{i<j}^{K}\prod_{k_i}^{N_i}\prod_{k_j}^{N_j}
\left(z_{k_i}^{(i)}-z_{k_j}^{(j)}\right)^{n_{ij}}
$$
takes into account correlations between the different components. Here,
$z_{k_i}^{(i)}$ denotes the complex position of the $k_i$-th particle of component $i$,
the exponents $m_j$ and $n_{ij}$ indicate the strength of the intra- and intercomponent
($i\neq j$) correlations, respectively. Similar generalizations have been suggested
by Qiu {\sl et al.} for multilayer QH systems \cite{macdon}
and by Morf in the framework of the 
FQHE hierarchy scheme \cite{morf}. We concentrate
on $K=4$, with $1=(\ua,+)$, $2=(\ua,-)$, $3=(\da,+)$, and $4=(\da,-)$, where
$\ua,\da$ denote the $z$-component of the physical spin and $\pm$ the
valley in graphene or the layer index in GaAs bilayer systems. 
We (artificially) break the spin-valley 
\footnote{In the following, we no longer distinguish between layer and valley index and
use the term {\sl valley} generically.}
symmetry and distinguish
between intracomponent pseudopotentials \cite{haldane} $V_m^A$, $V_m^{E-s}$
for different spin in the same valley, and $V_m^E$ for electrons in different valleys.
Notice that the wave function (\ref{eq01}) is an 
{\sl exact} ground state for a model interaction
$V_m^A>0$ for $m<\min(m_1,...,m_4)$, $V_m^E>0$ for
$m<\min(n_{12},n_{14},n_{23},n_{34})$, 
$V_m^{E-s}>0$ for $m<\min(n_{13},n_{24})$,
and $V_m^{A/E/E-s}=0$ otherwise. In GaAs bilayer QH systems, the difference between
inter- and intralayer interactions fixes $V_m^A=V_m^{E-s}$.

For general $K$, the component filling factors $\nu_j=\rho_j/n_B$ of $j$-particles
may be obtained from the usual 
zero-counting argument according to which the flux density $n_B$ equals
the sum of the $j$-component densities $\rho_j$ 
times the strength of their zero,
$n_B=\rho_j m_j +\sum_{i\neq j}\rho_i n_{ij}$,
for all $j$. With the help of the symmetric $K\times K$ exponent
matrix $M_K\equiv (n_{ij})$, where
$n_{ji}=n_{ij}$ and $n_{jj}\equiv m_j$, one may thus calculate 
the $\nu_j$ in a concise manner,
\beq\label{eq02}
\left(\nu_1, ..., \nu_K\right)^T=M_K^{-1}\left(1,...,1\right)^T
\eeq
if the matrix $M_K$ is invertible. 

\begin{center}
\begin{table}
\begin{tabular}{c|c|c|ccc}
$[m_1 m_2 m_3 m_4,n_e n_+ n_-]$ & $r$ & $\nu_T$ & $S_z$ & $I_z$ & $P_z$ \\
\hline\hline
$[3333,111]$ & $4$ & $2/3$ & $0$ & $0$ & $0$ \\
$[3333,033]$ & $2$ & $2/3$ & $-$ & $0$ & $-$ \\
\hline
$[3555,222]$ & $4$ & $2/5$ & $1/3$ & $1/3$ & $1/3$\\
$[3333,233]$ & $2$ & $2/5$ & $-$ & $0$ & $-$ \\
\hline
$[3535,222]$ & $4$ & $8/19$ & $0$ & $1/2$ & $0$ \\
\hline
$[5555,222]$ & $4$ & $4/11$ & $0$ & $0$ & $0$ \\
$[3737,233]$ & $3$ & $4/11$ & $-$ & $1/2$ & $-$ \\
$[3535,235]$ & $2$ & $4/11$ & $-$ & $1/2$ & $-$ \\
\hline
$[3333,333]$ & $1$ & $1/3$ & $-$ & $-$ & $-$ \\
\label{tab01}
\end{tabular}
\caption{\footnotesize{Examples of SU($4$) wave functions.}}
\end{table}
\end{center}

\begin{figure}
\centering
\includegraphics[width=6.5cm,angle=0]{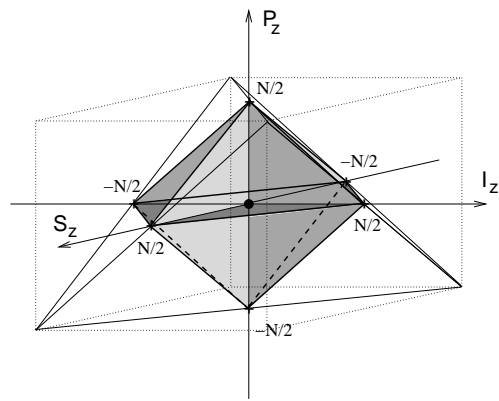}
\caption{\footnotesize{
The three polarizations, $S_z,I_z$, and $P_z$,
are restricted to the interior of 
a tretrahedron, completely explored in the case of a 
fully SU($4$)-symmetric state ($r=1$). 
For states with a matrix
rank $r=2$, the polarizations are restricted to quadratic planes with
fixed $S_z,I_z$ or $P_z$, e.g. $[3333,033]$ state with $I_z=0$ (plane in light gray). 
The $[3333,111]$ state, with $r=4$, has $S_z,I_z,P_z=0$ (black dot).
}}
\label{fig01}
\end{figure}

Tab. I shows examples of SU($4$) $N$-particle wave 
functions (\ref{eq01}), labeled by the set of exponents 
$[m_1 m_2 m_3 m_4,n_e n_+ n_-]$, where we have restricted the 10 exponents
to only 7, $n_+\equiv n_{13}$, $n_-\equiv n_{24}$, and $n_{e}\equiv
n_{12}=n_{14}=n_{23}=n_{34}$. This means that we treat all intervalley 
correlations on the same footing, with an exponent $n_e$, and the 
intravalley correlations for different spin are described by the exponent $n_+$ and 
$n_-$ in the valleys $+$ and $-$, respectively. 
Apart from the total filling 
factor $\nu_T=\nu_1+\nu_2+\nu_3+\nu_4$, the $[m_1 m_2 m_3 m_4, n_e n_+ n_-]$ wave
functions have the spin
$S_z/(N/2)=[\nu_1+\nu_2-(\nu_3+\nu_4)]/\nu_T$, the valley
$I_z/(N/2)=[\nu_1+\nu_3-(\nu_2+\nu_4)]/\nu_T$, and a third polarization
$P_z/(N/2)=[\nu_1+\nu_4-(\nu_2+\nu_3)]/\nu_T$.
These polarizations are good quantum numbers associated with
the mutually commuting
$z$-components of the SU($4$) spin, $\tau_z\otimes 1$, $1\otimes\tau_z$,
and $\tau_z\otimes\tau_z$, respectively, in terms of the diagonal Pauli matrix
$\tau_z$. Although they vary in the interval 
$[-N/2,N/2]$, they are not independent and indeed restricted to the interior
of a tetrahedron depicted in Fig. \ref{fig01}.

Some physical properties of the SU($4$) states may be characterized by
the rank $r$ of the matrix $M_4$.
In the case of
$r=4$, as e.g. for the $[3333,111]$ wave function at $\nu=2/3$ or
$[5555,222]$ at $\nu=4/11$, all $\nu_j$
are uniquely determined by Eq. (\ref{eq02}), and the $S_z,I_z$, and $P_z$ are thus
fixed. This is represented by the black dot in Fig. \ref{fig01}
for the $[3333,111]$ wave function. 
For $r=3$, three rows of $M_4$ are linearly independent. 
Consider the first row [$(\ua,+)$ component] to be a multiple of the third 
[$(\da,+)$ component].
In this example,
the wave function, such as $[3737,233]$ at $\nu=4/11$, represents a state with
a SU($2$) spin ferromagnet in the $+$ valley.  
Defining a combined filling factor for the $+$
component, $\nu_+=\nu_{\ua,+}+\nu_{\da,+}$, one may describe this state
alternatively by a SU($3$) wave function (\ref{eq01}), with an
invertable $M_3$ matrix. 
For $r=2$, one may e.g. realize a SU($3$) ferromagnet of three components,
whereas the combined three-component filling factor and that of the
fourth are fixed. Another possibility is that the
SU($2$) ferromagnet discussed
for $r=3$ is accompanied by another SU($2$) ferromagnet in the $-$ valley with
no coherence between the two of them. Examples are the $[3333,033]$ and
the $[3535,235]$ wave functions at $\nu=2/3$ and $4/11$, respectively. 
The polarizations are now restricted to planes with fixed $I_z$, as depicted in Fig.
\ref{fig01}. By a simple exchange of the components,
harmless in the case of an underlying SU($4$) symmetry, $S_z$ or
$P_z$ may play the role of $I_z$ so that the planes with fixed $I_z,S_z$, and
$P_z$ are equivalent. Both  
ferromagnetic states for $r=2$ may be described by a 
SU($2$) Halperin wave function with invertable matrix $M_2$. The
case $r=1$ represents a SU($4$) ferromagnet
the polarizations of which explore the full tetrahedron depicted in Fig.
\ref{fig01}. The only constraint is fixed by $\nu_T$, and one finds a
SU($4$) Laughlin wave function, where all possible states may be obtained
by any SU($4$) rotation of a state with $\nu_1=1/(2s+1)$ and 
$\nu_2=\nu_3=\nu_4=0$.

As becomes apparent from Tab. I, the same filling
factor may be realized by different wave functions with different matrix
rank $r$. This has to be contrasted to the SU($2$) case where one needs to invoke
the composite-fermion theory \cite{jain} to obtain competing states with 
different spin polarization at the same filling factor, such as e.g. at
$\nu=2/5$. However, filling factors which do not arise in U($1$) or SU($2$)
wave functions, such as e.g. $\nu=8/19$
(or $10/23$ and $26/47$, not shown in Tab. I), 
may be described by SU($4$) wave functions
with $r=4$. One may therefore principally expect a closer vicinity of FQHE states 
in SU($4$) than in SU($K<4$). 


In our numerical studies, we use an exact diagonalization on the sphere with a fully implemented SU($4$) invariance. All calculations are performed in the lowest LL. The various SU($4$) trial wave functions are obtained by tuning the pseudopotentials in the abovementioned manner. They appear as zero-energy ground states with the lowest number of flux quanta for a given interaction (similarly we obtain the Laughlin state or the Halperin states \cite{yoshioka}). 
The number of flux quanta threading the sphere is $2S$. 
Due to the large Hilbert spaces we need to consider, our calculations are restricted to the system sizes of $N=8$ fermions.

\begin{figure}
\centering
\includegraphics[width=3.5cm,angle=0]{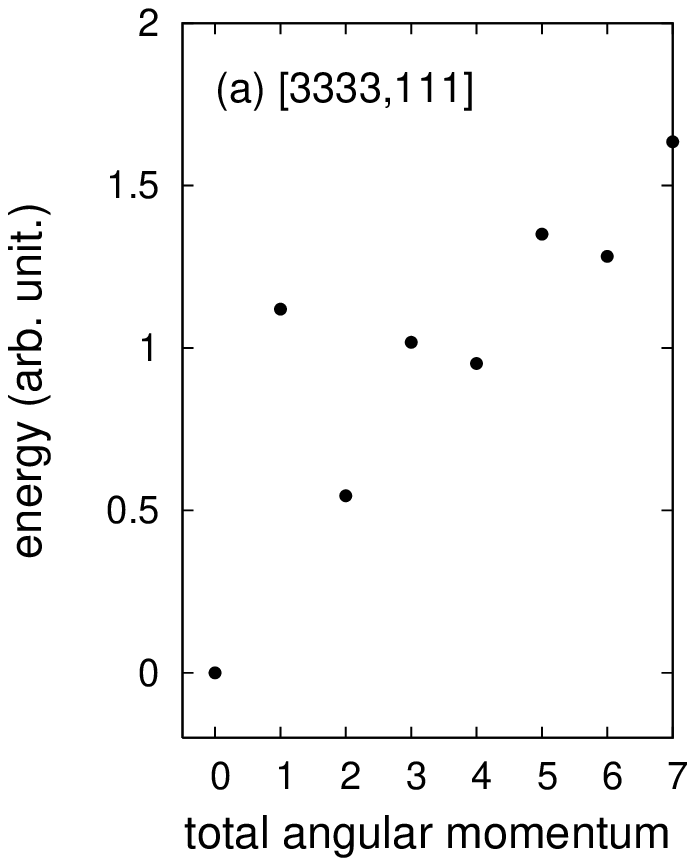}
\includegraphics[width=3.55cm,angle=0]{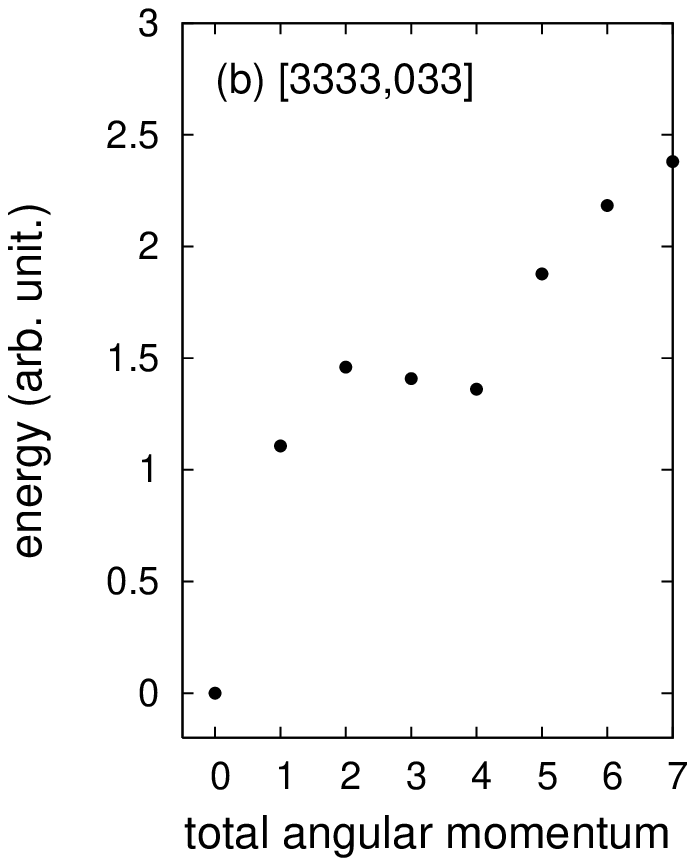}
\caption{\footnotesize{Low-energy spectra, for $N=8$ and $2S=9$, 
with contact potential in the LLL, built in such a way that 
the trial states (a) $[3333,111]$ and (b) $[3333,033]$ are the only 
ground states with zero energy. 
We only show the lowest-energy state for each value of the total angular
momentum $L$.
}}
\label{fig02}
\end{figure}

In the spherical geometry, the two states $[3333,111]$ and $[3333,033]$
occur at $2S=(3N/2)-3$.
Fig. \ref{fig02}(a) shows the energy spectrum of the FQHE at $\nu=2/3$, with $2S=9$ and $N=8$, for 
a choice of pseudopotentials $V_m^A=V_m^E=V_m^{E-s}=0$, except for 
$V_1^A=V_0^E=V_0^{E-s}=1$, 
for which $[3333,111]$ is expected to be the exact ground state. 
The ground state is found at a total angular momentum
$L=0$, which is, together with the finite gap to all excited states, a condition for an incompressible
FQHE state. Moreover, the ground state is non-degenerate and 
has $S_z=I_z=P_z=0$,
in agreement with the $[3333,111]$ state (Fig. \ref{fig01}). 
In Fig. \ref{fig02}(b), the energy spectrum ($2S=9$, $N=8$) is shown for a 
different pseudopotential choice 
with $V_m^A=V_m^E=V_m^{E-s}=0$, except for $V_0^{E-s}=V_1^{E-s}=V_1^A=1$ -- 
a model for which 
the $[3333,033]$ state is expected to be the exact ground state. 
One finds a gapped $L=0$ ground state with $I_z=0$ and a degeneracy
due to a free choice of $S_z$ and $P_z$ within the light gray plane in 
Fig. \ref{fig01}, as for a $[3333,033]$ state.

\begin{figure}
\centering
\includegraphics[width=3.5cm,angle=0]{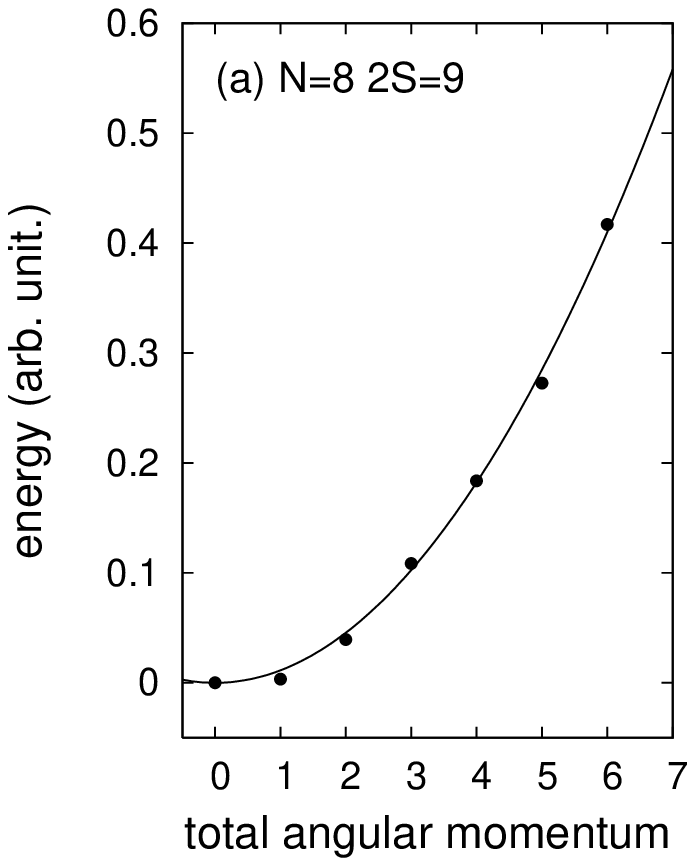}
\includegraphics[width=3.5cm,angle=0]{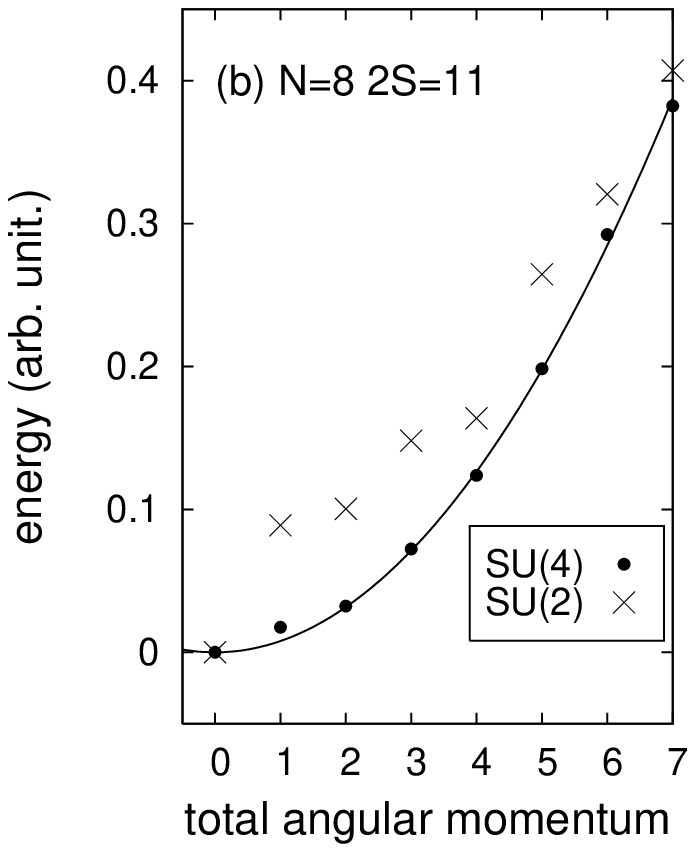}
\caption{\footnotesize{Low-energy spectra with Coulomb
interaction in the lowest LL {\sl (a)} for $2S=9$ and $N=8$; {\sl (b)} 
for $2S=11$ and $N=8$. We show only the lowest energy for each value of
$L$ in {\sl (a)}, but also the spectrum for $S_z=0,I_z=0,
~P_z=4$ (crosses), for comparison with the SU(2) case \cite{AC}.
The continuous line shows a 
$L^2$ fit through the 
lowest-energy states. The only fit parameter is known with a precision of 
$1\%$ (if removing 
the $L=1$ point) in both cases.
}}
\label{fig03}
\end{figure}

The Coulomb interaction in the lowest (central) LL turns out to be more involved. 
We have first performed Monte-Carlo calculations for up to $N=70$ particles, 
which indicate that in this case the $[3333,111]$ state is
much lower in energy than the $[3333,033]$ state when compared to the statistical error for
both states. 
Fig. \ref{fig03}(a) shows the energy spectrum for $2S=9$ and $N=8$, obtained from exact diagonalization
of the lowest-LL Coulomb interaction. The
spectrum is reminiscent of that for $2S=11$ and $N=8$ [Fig. \ref{fig03}(b)],
but it is clearly different from those of the
$[3333,111]$ and $[3333,033]$ states. Both spectra show a $L=0$ ground state that is connected to 
a collective mode with a striking $L^2$ behavior (black lines). However the ground-state degeneracy
is different, as shown in Tab. II. For $2S=11$, we obtain a $[N/2,N/2]=[4,4]$ multiplet \cite{quesne},
in agreement with results by T\"oke and Jain \cite{toke2}. The non-degenerate $S_z=I_z=0,~P_z=4$ state,
is, up to an interchange of $S_z$ and $P_z$, the state described by Apalkov and Chakraborty \cite{AC},
who have considered an internal SU($2$) valley symmetry and a complete spin polarization. Indeed, the 
spectra coincide, but Fig. \ref{fig03}(b) shows clearly that this is only the top of the iceberg and that states with different SU(4) polarization also
belong to the ground-state manifold. More importantly, the low-energy 
properties are dominated by a $L^2$ mode (Fig. \ref{fig03}) we
conjecture to be associated with collective 
spinlike waves the symmetry of which is
determined by a subgroup of SU(4). Notice that such mode would not affect the 
incompressibility of the ground state.

\begin{center}
\begin{table}
{\footnotesize
\begin{tabular}{c||c|c|c|c|c|c|c}
$(S_z I_z P_z)$ & $(0 0 0)$ & $(0 0 2)$ & $(0 0 4)$ & $(0 1 1)$ & $(0 1 3)$ & $(0 2 2)$ & $(1 1 2)$\\
\hline
deg. $(2S=11)$ & 3 & 2 & 1 & 2 & 1 & 1 & 1 \\
\hline
deg. $(2S=9)$ & 3 & 1 & $-$ & 3 & $-$ & 1 & 1 \\
\end{tabular}}
\caption{\footnotesize 
{Degeneracy of the ground state 
for the Coulomb interaction in the different $0\leq S_z \leq I_z \leq P_z$ 
sectors for $N=8, 2S=11$ 
(second row) and $N=8, 2S=9$ (third row). $(004)$ and $(013)$ 
are not ground states for $2S=9$.}
}
\label{degeneracies}
\end{table}
\end{center}

\begin{figure}
\centering
\includegraphics[width=8cm,angle=0]{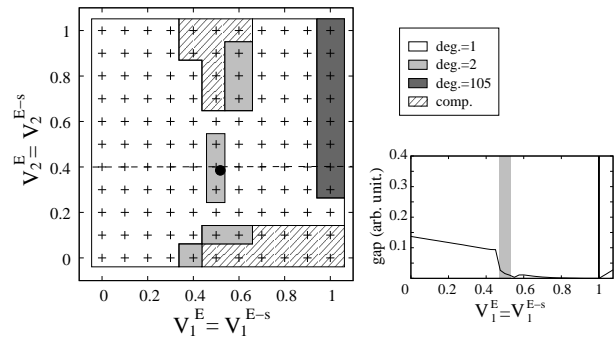}
\caption{\footnotesize{Left panel : schematic view of the ground state degeneracy for pseudopotentials $V^E_1$ and $V^E_2$ varying 
from $0$ to $1$, evaluated on a regular grid ($11 \times 11$ values),
for $2S=9,N=8$. The white region corresponds to a non-degenerate 
ground state, light grey regions are doubly degenerate ones with $S_z=I_z=P_z=0$ and dark grey regions are 105$\times$ degenerate 
 ground states. Hatched regions are compressible ground states with $L\neq0$. The black dot corresponds to the ratios $V^E_1=V_1^{C}/V_0^{C}$ and $V^E_2=V_2^{C}/V_0^{C}$, where $V_m^{C,0}$ are Coulomb pseudopotentials in the lowest LL. Right panel : gap as a function of $V_1^E$ for fixed $V_2^E=0.4$. The peak corresponds to the 105$\times$ degenerate ground state. For $V_1^E > 1$, there is a 6 times degenerate ground state with one of the quantum number maximaly polarized and the others unpolarized.}
}
\label{fig04}
\end{figure}

In order to make a connection between the $[3333,111]$ ground state and that obtained for 
Coulomb interaction in the lowest LL for $2S=9$ and $N=8$,
we evaluate the ground states when varying the
pseudopotentials $V_1^E$ and $V_2^E$ from $0$ (exact model for the $[3333,111]$ state) to
$1$, keeping $V_1^A=V_0^E=1$. Here, we set $V_m^E=V_m^{E-s}$ for all $m$.
Fig. \ref{fig04} shows the resulting phase diagram, in terms of the ground state degeneracies. 
The $[3333,111]$ state is stable over a large region in the phase diagram, but one also obtains
doubly degenerate (at moderate $V_1^E\sim 0.4 ... 0.5$) and compressible ground states with 
$L\neq 0$. Most saliently, a 105$\times$ degenerate ground state is obtained at 
$V_1^E=1$, for $V_2^E\gtrsim 0.25$, such as that corresponding to the Coulomb
interaction (dashed line in Fig. \ref{fig04}). This degeneracy is precisely that of
a $[4,4]$ SU($4$) multiplet. The phase is critical in the sense that it is
destroyed as soon as we slightly deviate from $V_0^E=V_1^E=V_1^A=1$.

In conclusion,
we have analyzed a SU($4$) generalization of Halperin's wave function \cite{halperin}, which
may be a promissing approach towards a SU($4$) FQHE eventually occuring in high-mobility
graphene sheets. These trial wave functions yield incompressible states
at filling factors, which may not, in some instances, be described by Laughlin \cite{laughlin},
SU($2$) Halperin \cite{halperin}, or SU($4$) composite-fermion wave functions \cite{toke2}, such
as at $\nu=8/19$. 
Moreover, the rank of the exponent matrix may be related to ferromagnetic
properties of the trial states within the SU(4) symmetry.
Whether these FQHE states are realized, depends on the precise form of
the interaction potential. At $\nu=2/3$, two candidates, $[3333,111]$ and
$[3333,033]$, with different residual symmetry, yield incompressible FQHE states for
appropriately chosen pseudopotentials. The Coulomb interaction, however, chooses a state
with a pronounced $L^2$ collective (Goldstone) mode, {\sl both} for $2S=9$ and $11$. This 
mode is thus likely to survive in the thermodynamic limit. 

We would like to thank B. Dou\c cot, P. Lederer, R. Moessner, R. Morf, and C. T\"oke for 
stimulating discussions. MOG acknowledges financial funding from the Agence 
Nationale de Recherche (ANR-06-NANO-019-03).

\end{document}